\documentclass[showpacs,preprintnumbers,amsmath,amssymb,twocolumn,superscriptaddress,prb]{revtex4}
\usepackage{graphicx}
\usepackage{amsfonts}
\usepackage{dsfont}
\usepackage{amsmath, amsthm, amssymb}

\newcommand{\Mso}{M_{\mathrm{s}}}
\newcommand{\chimso}{\chi_{\mathrm{s}}}

\def\be{\begin{equation}}
\def\ee{\end{equation}}
\def\bea{\begin{eqnarray}}
\def\eea{\end{eqnarray}}

\begin{document}
\title[Thermodynamic Properties near the onset of Loop-Current Order in high-$T_c$ superconducting cuprates]
{Thermodynamic Properties Near the Onset of Loop-Current Order  in high-$T_c$ Superconducting Cuprates}
\author{M. S. Gr{\o}nsleth}
\affiliation{Department of Physics, Norwegian University of
Science and Technology, N-7491 Trondheim, Norway}
\author{T. B. Nilssen}
\affiliation{Department of Physics, Norwegian University of
Science and Technology, N-7491 Trondheim, Norway}
\author{E. K. Dahl}
\affiliation{Department of Physics, Norwegian University of
Science and Technology, N-7491 Trondheim, Norway}
\author{E. B. Stiansen}
\affiliation{Department of Physics, Norwegian University of
Science and Technology, N-7491 Trondheim, Norway}
\author{C. M. Varma}
\affiliation{Department of Physics and Astronomy, University of California, Riverside, CA 92521, USA}
\author{A. Sudb{\o}}
\affiliation{Department of Physics, Norwegian University of
Science and Technology, N-7491 Trondheim, Norway}
\date{Received \today}
\begin{abstract}
We have performed large-scale Monte Carlo simulations on a two-dimensional generalized Ashkin-Teller
model to calculate the thermodynamic properties in the critical region near its transitions. The Ashkin-Teller 
model has a pair of Ising spins at each site which interact with neighboring spins through pair-wise and 4-spin 
interactions. The model represents  the interactions between orbital current loops in $Cu O_2$-plaquettes of 
high-$T_c$ cuprates, which order with a staggered magnetization $\Mso$ inside each unit-cell in the underdoped 
region of the phase diagram below a temperature $T^*(x)$ which depends on doping. The pair of Ising spins per 
unit-cell represent the directions of the currents in the links of the current loops. The generalizations are 
the inclusion of anisotropy in the pair-wise nearest neighbor current-current couplings consistent with the 
symmetries of a square lattice and the next nearest neighbor pair-wise couplings. We use the Binder cumulant to 
estimate the correlation length exponent $\nu$ and the order parameter exponent $\beta$. Our principal results 
are that in a range of parameters, the Ashkin-Teller model as well as its generalization has an order parameter 
susceptibility which diverges as $T \to T^*$ and an order parameter below $T^*$. Importantly, however, there 
is no divergence in the specific heat. This puts the properties of the model in accord with the experimental 
results in the underdoped cuprates. We also calculate the magnitude of the "bump" in the specific heat in the 
critical region to put limits on its observability. Finally, we show that the staggered magnetization couples 
to the uniform magnetization  $M_0$ such that the latter has a weak singularity at $T^*$ and also displays a 
wide critical region, also in accord with recent experiments. 
\end{abstract}
\maketitle
\section{Introduction}\label{Introduction}
It has been proposed \cite{Varma1, Varma2} that the properties of the cuprate compounds are controlled by the 
onset of a  time-reversal and inversion violating order parameter below a temperature $T=T^*(x)$, which depends 
on the doping $x$.  $T^*(x) \to 0$ for $x\to x_c$ in the superconducting range of compositions, thus defining 
a quantum critical point. The quantum critical fluctuations associated with the  breakup of the specific 
order proposed have been shown \cite{Aji-Varma} to be of the scale-invariant form hypothesized to lead to  
a Marginal Fermi Liquid \cite{MFL_Varma}, which explains the anomalous transport properties of these 
compounds. $T^*(x)$ is identified with the observed onset of the pseudogap properties in the cuprates. 
\par
\begin{figure}[htbp]
  \centerline{\hbox{\includegraphics[width=50mm]{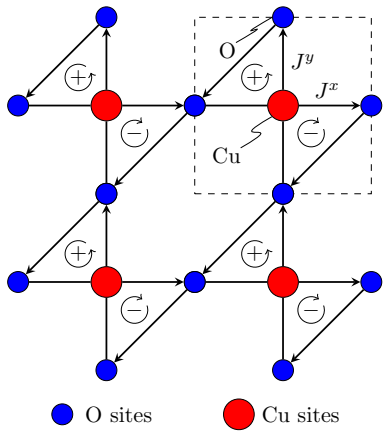}}}
  \caption{(Color online) The circulating current phase $\Theta_{II}$  \cite{Varma2}. 
  The $Cu$ sites are red  circles, $O$ sites are blue.  The unit 
  cell is shown by the dashed square. A staggered magnetic moment pattern 
  within each unit cell that repeats from unit cell to unit cell (the curl 
  of the directed circles) is indicated. The currents $J^x$ and $J^y$
  represent the horizontal end vertical currents, respectively, to be used 
  in the derived effective model, Eq. \ref{Model} 
  below. Physically, they represent the {\it coherent parts} of the orbital fermionic 
  currents in the problem.}
\label{orbital_currents}
\end{figure}

A major difficulty in accepting these ideas is that there is no observed specific heat divergence near $T^*(x)$ 
in any cuprate. On the other hand, there now exists significant evidence for long-range order with  a spatial 
symmetry consistent with orbital currents of the form shown in Fig.~\ref{orbital_currents} in three different  
families of cuprates \cite{Fauque,Mook, Greven, Kaminski} which have been investigated so far.  There is 
also evidence of a weak singularity at $T^*(x)$ in the uniform magnetic susceptibility \cite{Leridon}. 
\par
In view of this situation, it is important to investigate whether or not the proposed models for 
these novel broken symmetries are consistent simultaneously with long-range order without an observable 
signal in the specific heat in the measurements made hitherto, and also whether it does give rise to 
observable features in the uniform magnetization induced by an external magnetic field \cite{Leridon}. 
\par
The particular form of proposed hidden order is one of spontaneously generated fluxes in the $O$-$Cu$-$O$
plaquettes of the $CuO_2$ unit cell such that currents flow in two oppositely directed loops in each unit-cell, 
as depicted for one of the four possible domains in Fig. \ref{orbital_currents}. (See also Fig. 1 of  Refs. 
\onlinecite{Varma2,Borkje_Sudbo}). The {\it staggered} orbital magnetic moments within each $CuO_2$ unit cell 
repeats from unit cell to unit cell so that the translational symmetry of the lattice remains unaltered. These 
circulating current patterns are generated by a nearest-neighbor repulsion $V$ between $Cu$ and $O$-atoms in 
the $CuO_2$-sheets. The effect of such a repulsive $V$-term has been extensively investigated in $1D$ $CuO$-chains, 
where it has been shown to drive charge-transfer instabilities and superconductivity 
\cite{1DCuO-works1,1DCuO-works2,1DCuO-works3}. In  Ref. \onlinecite{Greiter}, the existence of current-loop 
ordering was not confirmed, but the work was carried out on a truncated effective $t-J$ model of $8$ $Cu$ sites. 
Moreover, the ground state was spin-polarised with finite momentum, which would not be representative of the 
large-scale physics of interest in the system. The truncation of the Hilbert space used in Ref. \onlinecite{Greiter} 
furthermore requires so large values of onsite Coulomb repulsion on oxygen sites that it is probably outside the 
parameter regime of the high-$T_c$ cuprates \cite{Hybertsen}. This motivated the authors of Ref. \onlinecite{Weber} 
to undertake a large-scale study of the issue of current-loop ordering on much larger systems using the full 
three-band model of the $CuO_2$ planes via variational Monte Carlo simulations. These authors find clear evidence 
for current-loop ordering. Other types of current-patterns and charge-fluctuations are also possible 
\cite{Varma1, Lee_Choi_2001, Chakravarty-ddw}. 

\par

\section{Fundamentals}\label{fund}
In this section, we  present the effective model of fluctuating orbital currents we 
study in this paper, along with the definitions of the thermodynamic quantities we  
compute, as well as some remarks on the critical exponents of the problem with
emphasis on the particular status of the specific heat exponent of the problem
at hand.  

\subsection{Model}\label{model_desc}
The  effective model we perform Monte Carlo simulations on, has been derived from a microscopic 
description of the $CuO_2$-planes of high-$T_c$ cuprates elsewhere \cite{Borkje_Sudbo,Borkje_Thesis}. 
It turns out to be a generalization of the model initially proposed to describe the statistical 
mechanics of loop current order \cite{Varma2,Aji-Varma} in which some terms  allowed by symmetry 
were omitted. The action $S$ is written on the form $S = S_{\rm{C}} + S_{\rm{Q}}$, where $S_{\rm{C}}$ 
is the classical piece of the action, and $S_{\rm{Q}}$ is part of the action that is needed in the 
quantum domain of the theory. In this paper, we will focus on discussing the effects of thermal 
fluctuations, and we will therefore not need $S_{\rm{Q}}$. The classical part of the action, 
$S_{\rm{C}}$, is given by \cite{Borkje_Sudbo,Borkje_Thesis}  
\begin{eqnarray}
\label{Model}
S_{\rm{C}} &=&  -  \beta  \sum_{\langle {\bf r}, {\bf r'}\rangle} \left( K_x ~ J^{x}_{\bf r} J^{x}_{\bf r'}
                 + K_y ~ J^{y}_{\bf r} J^{y}_{\bf r'}  \right) \nonumber \\
               & -  & \beta  \sum_{\langle \langle {\bf r} ,{\bf r'} \rangle \rangle} 
                   K^{xy}_{{\bf r}{\bf r}'}  ~  \Big( J^x_{\bf r} J^y_{\bf r'} + J^y_{\bf r} J^x_{\bf r'} \Big) \nonumber
                  \\
              & - &  \beta ~K_4  \sum_{\langle {\bf r}, {\bf r'}\rangle} 
  ~ J^{x}_{\bf r} J^{x}_{\bf r'}  J^{y}_{\bf r} J^{y}_{\bf r'}.  
\end{eqnarray}
Here, $\langle {\bf r},{\bf r'} \rangle$ and $\langle \langle {\bf r},{\bf r'} \rangle \rangle$ 
denote nearest-neighbor and next-nearest-neighbor summations, respectively. $\beta = 1/T$ where $T$ 
is temperature, and we work in units where Boltzmann's constant $k_B=1$. We will only consider the 
directions $\pm$ of the current variables $J^{x,y}$, assuming as in other similar two-dimensional 
models that their amplitudes are smoothly varying with temperature and do not determine the critical 
properties. Note that there is always also a current in the $O-O$ links whose magnitude is equal to 
that of $J^x$ which has the same magnitude as $J_y$. Therefore, no current flows out of any $O-Cu-O$ 
triangular plaquette. Due to the restriction that no current flows out   of any $O-Cu-O$ plaquette, 
there is no need to specify the $O-O$ currents. The variables $J^x,J^y$ are then the same as  the 
$\sigma = \pm1$ and $\tau = \pm1$ Ising variables introduced earlier \cite{Aji-Varma}. Fluctuations 
$(J^x_{\bf r} \to - J^x_{\bf r}, J^y_{\bf r} \to  J^y_{\bf r})$ corresponds to going from the depicted 
current pattern (Fig. \ref{orbital_currents}) to a new one which is obtained by a counterclockwise rotation 
by $\pi/2$, $(J^x_{\bf r} \to J^x_{\bf r}, J^y_{\bf r} \to  -J^y_{\bf r})$ corresponds to clockwise 
rotation of $\pi/2$, and $(J^x_{\bf r} \to -J^x_{\bf r}, J^y_{\bf r} \to  -J^y_{\bf r})$ to a rotation 
of $\pi$. 
\par
If one ignores the next-nearest neighbor terms and takes $K_x = K_y$, one gets the Ashkin-Teller (AT) 
model \cite{AT}, for which several exact results are known \cite{Baxter} asymptotically close to the 
phase transition lines. However, since the currents are bond-variables, one necessarily has an 
anisotropy in the nearest neighbor interactions \cite{Borkje_Sudbo,Borkje_Thesis}, such that for 
${\bf r - r'} = \pm {\bf \hat{x}}$, $K_x = K_l$ and $K_y = K_t$, whereas when 
${\bf r - r'} = \pm {\bf \hat{y}}$, $K_x = K_t$ and $K_y = K_l$. It is important to investigate whether 
this anisotropy is an {\it irrelevant} perturbation. We will in the following denote  the anisotropy 
by the parameter  $A \equiv K_t/K_l$. Similarly, it is interesting to investigate the effect of 
the next-nearest neighbor interaction  given by the parameter $K^{xy}_{{\bf r}{\bf r}'} = K^{xy}$ 
when ${\bf r - r'} = \pm ({\bf \hat{x}} + {\bf \hat{y}})$ and $K^{xy}_{{\bf r}{\bf r}'} = - K^{xy}$ 
when ${\bf r - r'} = \pm ({\bf \hat{x}} - {\bf \hat{y}})$.   
\par
Let us comment briefly on the terms appearing to quartic order, most of which either are constants or 
renormalize the quadratic piece of the action. Note that four Ising variables of two distinct species 
all located on one single lattice site, simply contribute a constant to the action. If we now limit 
ourselves to terms that have four $J$-fields distributed on two nearest-neighbor lattice sites, only 
two distinct possibilities exist. Firstly, we may  have a term with three $J$'s on one lattice site and 
one $J$ on a nearest-neighbor site. This merely represents a renormalization of the quadratic couplings. 
Secondly, we may have two $J$'s on one lattice site and another two on a nearest neighbor lattice site. 
Unless there are two distinct species of $J$'s on each of the lattice sites, such a term will represent a 
constant contribution to the action. If the $J$'s on each lattice site are of distinct species, the term 
will be of the AT-form, as written above. We will ignore terms that have $J$-fields distributed on three 
or four distinct lattice sites, such as for instance plaquette terms, as these are generated by much 
higher order terms \cite{Borkje_Sudbo,Borkje_Thesis}. 

\subsection{Thermodynamic quantities}\label{therm_quant}
In this paper, we calculate  the evolution of the specific heat, the staggered orbital magnetic moment 
as well as the susceptibility of the staggered orbital magnetic moment as we vary $K_4$  in Eq. \ref{Model}.  
We also perform finite-size scaling on the magnetization and the Binder cumulant (see below). The specific 
heat $C_v$ is given by
\begin{eqnarray}
\label{C_v}
C_v = \frac{1}{L^2} 
\langle  ({\cal S}_{\rm{C}}  - \langle  {\cal S}_{\rm{C}} \rangle)^2 \rangle.
\end{eqnarray} 
\begin{figure}[htbp]
\centerline{\hbox{\includegraphics[width=60mm]{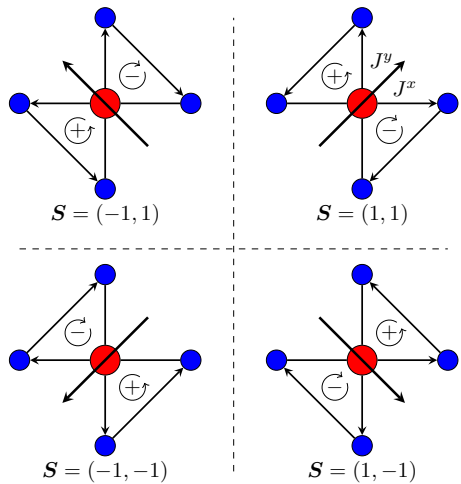}}}
\caption{(Color online) An illustration of the  pseudo-``spin''
${\bf S} = (J^x_{{\bf r}}, J^y_{{\bf r}})$ we use to compute the staggered 
order parameter and its susceptibility, Eqs. \ref{stag_mag} and \ref{chi_mso}. }
\label{pseudospin}
\end{figure}
Considering Fig. \ref{pseudospin}, we see that we may define a pseudo-``spin" {\bf S} on each lattice given by 
${\bf S}_{{\bf r}} \equiv (J^x_{{\bf r}},J^y_{{\bf r}})$. The various states of the system 
are then described by a $4$-state clock pseudospin  ${\bf S}_{{\bf r}} = (\pm 1, \pm 1)$ on 
a $2$-dimensional square lattice. We define the staggered order parameter in the standard 
way it would be defined for a clock model, namely  
\begin{eqnarray}
\langle \Mso \rangle \equiv \frac{1}{L^2} ~\left\langle \sqrt{ \frac{(m^x)^2  +  (m^y)^2 }{2} } \right\rangle,
\label{stag_mag}
\end{eqnarray}
where $m^\alpha \equiv  \sum_{{\bf r}} J^\alpha_{{\bf r}}, \alpha \in (x,y)$.
The susceptibility of this staggered order parameter is given by
\begin{eqnarray}
\label{chi_mso}
\chimso  =  
\frac{1}{2L^2 T}
\left[ \langle   (m^x)^2+(m^y)^2 \rangle  - \langle  \sqrt{(m^x)^2+(m^y)^2}  \rangle^2 \right].
\end{eqnarray}
We will contrast the singularities in these quantities with the evolution of the anomaly in 
the specific heat as the parameter $K_4$ is varied. While the above staggered moment does not 
couple linearly to an external uniform magnetic field, it couples to a field-induced uniform 
magnetic moment via a quartic term in the free energy. The field-induced uniform magnetization 
must therefore have a non-analytic behavior across the phase transition where the staggered 
magnetization associated with the ordering of the orbital currents sets in. We will return 
to this point in Section \ref{uniform-chi}. 
\par
For the purposes of extracting the critical exponent $\nu$, we consider the Binder cumulant, 
defined by
\begin{eqnarray}
\label{Binder}
G \equiv \frac{\langle  m^4 \rangle}{\langle m^2 \rangle^2},
\end{eqnarray}
where $m^2 = (m^x)^2 + (m^y)^2$, corresponding to the magnetization order parameter
$\langle |{\bf m} |\rangle$, whose critical exponent $\beta$ is given in Eq. \ref{nubeta}
for the AT-model \cite{Baxter}. In the ordered phase, $G=1$. For an $N$-component order parameter, 
$G=(N+2)/N$ in the disordered phase. In our case, therefore, $G$ will exhibit a rise from $1$ to 
$2$ as the systems disorders. When computing this quantity for different $L$ and plotting it as a 
function of $T$, the curves should in principle cross at the same point, thus defining $T_c$. On 
the other hand, plotting it as a function of $L^{1/\nu} |(T-T_c)/T_c)|$, all the curves will collapse 
on top of each other. By adjusting $\nu$ to get data-collapse, one obtains the correlation length 
exponent. Furthermore, the order-parameter exponent $\beta$ is obtained from the magnetization 
$\Mso$ for various system sizes by considering the quantity $L^{\beta/\nu} \Mso$ and adjusting 
$\beta$ and $\nu$ so as to obtain data-collapse when plotting this quantity as a function of 
$L^{1/\nu} |(T-T_c)/T_c)|$.  

\subsection{Critical exponents}\label{crit_exp}
Note that although the $K_x$ and $K_y$ couplings between the two different types of Ising fields 
in this model are anisotropic \cite{note-anisotropy,Borkje_Sudbo,Borkje_Thesis}, there is only 
one (doubly degenerate Ising) phase transition in the system for $K^{xy}=0; K_4 = 0$. Hence, as 
the four-spin coupling $K_4$ is changed from $0$, the Ising critical point evolves into a single 
phase-transition line with non-universal critical exponents \cite{Baxter}. In particular, the 
specific heat exponent $\alpha$ becomes negative, with the transition line itself being a selfdual 
critical line \cite{Baxter}. In this sense, the model is similar to an {\it isotropic} AT model, 
where the exact result for the  critical exponents are known, and  given by  \cite{Baxter}  
\begin{eqnarray}
\alpha = \frac{2-2y}{3-2y}; ~~ \beta = \frac{1}{8} ~ \left( \frac{2-y}{3-2y} \right).
\label{nubeta} 
\end{eqnarray}
From this, we deduce the susceptibility exponent $\gamma = 14 \beta$ and the correlation 
length exponent $\nu = 8 \beta$ from standard scaling relations. Note that the ratios 
$\gamma/\nu = 7/4$  and $\beta/\nu = 1/8$ are universal and independent of $y$. (It is 
also interesting to note that the anomalous scaling dimension $\eta=1/4$ and the magnetic
field exponent $\delta=15$, precisely as in the $2D$ Ising model). Here $y= 2 \mu/\pi$ and 
$\cos(\mu) = [e^{4 K_4/T_c}-1]/2$ \cite{Baxter}. Hence, for $K_4 \leq 0$, we have 
$\pi/2 \leq  \mu < 2 \pi/3$, such that $ 1 \leq y < 4/3$. 
\par
\begin{figure}[htbp]
  \centerline{\hbox{\includegraphics[width=75mm]{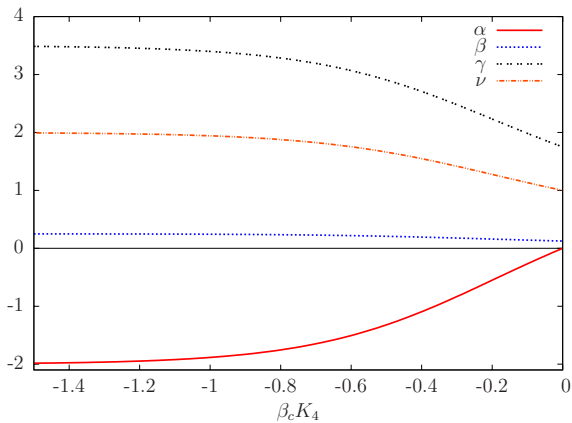}}}
  \caption{(Color online) 
Critical exponents $\alpha$, $\beta$, and $\gamma$ from the Ashkin-Teller model, as a function 
of the four-spin coupling $\beta_c K_4 \leq 0$ \cite{Baxter}. In this parameter range, we have 
$-2 < \alpha \leq 0$, $ 1/8 \leq \beta < 1/4$,   $7/4 \leq \gamma < 7/2$, and $1 \leq \nu < 2$.}
\label{exponents}
\end{figure}

These exponents are plotted in Fig. \ref{exponents}. The most extreme deviation from the $2D$ Ising 
values $\alpha=0, \beta=1/8, \gamma=7/4, \nu=1$ is given by the case $K_4 \to -\infty, y=4/3$, where 
$\alpha=-2, \beta =1/4, \gamma=7/2, \nu=2$.  Note the {\it increase} of $\gamma$ and $\nu$, (which 
implies a weak increase in $\beta$ for increasing $-K_4$ due to the proportionality factors $14$ and 
$8$ given below Eq. \ref{nubeta}), while we have a {\it substantial reduction} of $\alpha$ to 
{\it negative} values as $-K_4$ increases. This is traceable to the numerator $2-2y$ in $\alpha$ compared 
to the numerator $2-y$ in $\beta$, $\gamma$, and $\nu$ (while $\eta$ and $\delta$ are independent of $y$). 
Hence, the specific heat exponent stands out as very special in the model Eq. \ref{Model}. This fact is 
by far the single most dramatic difference between the critical behavior of Eq. \ref{Model} and the $2D$ 
Ising model. The $K_4$-term with $K_4 < 0$ simultaneously  {\it suppresses} singularities in the specific 
heat, and {\it enhances} singularities both in the susceptibility corresponding to the staggered orbital 
magnetization of Fig. \ref{orbital_currents} and in the one associated with a field-induced uniform 
magnetization (see Section \ref{uniform-chi}). 
\par

\section{Monte Carlo results}\label{MC:results}
The Monte Carlo computations were performed using the standard single-spin update Metropolis-Hastings algorithm 
\cite{Metropolis,Hastings}, making local updates of the Ising-fields $J^{x}_{{\bf r}}$ and $J^{y}_{{\bf r}}$, as 
well as local updates of the composite Ising-field $J^{x}_{{\bf r}} J^{y}_{{\bf r}}$ at each lattice site. The 
system-grid is defined by two $2$-dimensional subgrids, one for each Ising-field, and the local updates were 
performed  for all points on the grid. All the Ising-fields on both subgrids were initially set to 1. We started 
all simulations at the high-temperature end, and discarded the first $100000$  sweeps for the purpose of initially 
thermalizing the system. After that, measurements were made for every $100$ sweeps. The system sizes that were 
considered were $L \times L$ with $L=64,128,256,512$. 
For each value of $T$, we ran up 
to $3 \cdot 10^6$ MC sweeps for $L=64,128,256$ and sampled the system
for every $100$ MC sweeps over the lattice, while we used $5 \cdot 10^6$ MC sweeps for $L=512$
and sampled the system for every $150$ MC sweeps over the lattice.
We have 
checked that satisfactory convergence is well established by the time we get to system sizes of $L=512$, and we 
therefore largely present results for these largest systems only, apart from Fig. \ref{multi_cv} and the finite-size 
scaling results that will be presented for the Binder-cumulant(see below). In all simulations, we have set $K_l=1.0$, 
such that all other couplings are measured relative to this parameter. In these units, the critical temperature $T_c$ 
of the system for $A=1.0, K^{xy}=0, K_4=0$ is given by $T_c = 2/\ln(1+\sqrt{2}) \approx 2.27$. This sets the scale 
of the critical temperatures in the plots we will show below.   

\subsection{Specific heat}\label{Cv}
\begin{figure}[htbp]
\centerline{\hbox{\includegraphics[width=80mm]{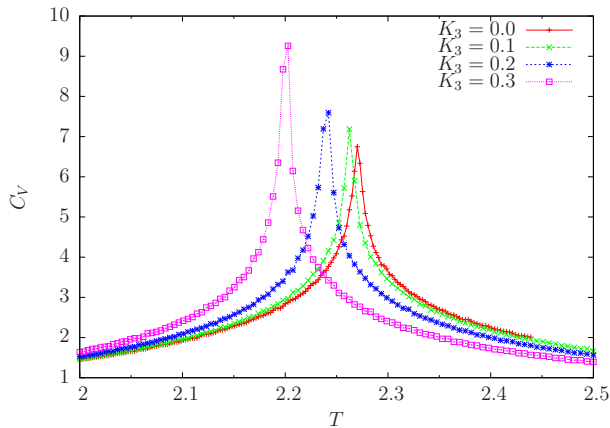}}}
\caption{(Color online) Specific heat as a function of  temperature $T$ for the classical part of 
the model in  Eq. \ref{Model}, with $A=1.0$ and $K_4=0.0$, for various values of $K^{xy} = 0.0,0.1,0.2,0.3$, 
and system size $L=512$. The amplitude of the logarithmic specific heat of the Ising model ($K^{xy} = 0$), 
is enhanced as  $K^{xy}$ increases, but the anomaly remains logarithmic. The critical temperature of the 
$2D$ pure Ising model is given by $T_c = 2/\ln(1+\sqrt{2}) \approx 2.27$ in units where Boltzmanns constant 
$k_B=1$. Note also that for this set of parameters, $K^{xy}$ hardly alters $T_c$ of the model with $K^{xy}=0$.}
\label{CV_fig0}
\end{figure}

Let us first investigate what effect $K^{xy}$ has on the logarithmic singularity of the $2D$ Ising model. In Fig. 
\ref{CV_fig0}, we show the specific heat  for $A=1.0$ and $K_4=0$, upon varying $K^{xy}=0.0,0.1,0.2,0.3$. We have  
limited the variations in $K^{xy}$ because it can be shown in mean-field calculations that the order parameter 
changes the translational symmetry for large enough $K^{xy}$ and a diagonal "striped" order is favored. It is seen 
that the $K^{xy}$ term in this parameter range leaves the logarithmic singularity of the anisotropic 
double-Ising model (Eq. \ref{Model} with $K^{xy}=0, K_4=0$) unaltered, only the amplitude of the singularity is changed. 

\par
We now investigate the effect of four-spin interactions $\propto K_4$. We will only consider negative 
values of $K_4$ in this paper. Then the four-spin term tends to promote a non-uniform ground state with 
antiferromagnetic ordering in the composite variable $J^{x}_{\bf r}  J^{y}_{\bf r}$, thus frustrating 
the Ising terms in Eq. \ref{Model}. It is known from the phase diagram of the AT model \cite{AT} that the 
ordered phase has a different symmetry in the regions $-1<K_4/K_l <1, K_4/K_l <-1$ and $K_4/K_l >1$. The 
region of special interest is $-1<K_4/K_l<0$ in which the AT model has a self-dual line of critical points 
\cite{Jose-Kadanoff,Kadanoff-Brown}. This is consistent with the microscopic model, which may exhibit a 
negative sign of the four-spin interaction term.
\par
\begin{figure}[htbp]
\centerline{\hbox{\includegraphics[width=80mm]{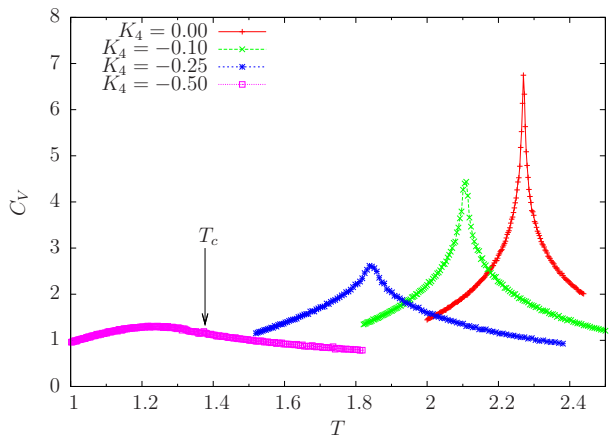}}}
\caption{(Color online) Specific heat as a function of temperature $T$ for the classical part of the 
generalized AT model Eq. \ref{Model}, with $A=1.0$ and $K^{xy}=0.0$, for various values of 
$K_4 = 0.0,-0.1,-0.25,-0.5$, and system size $L=512$. The vertical scale is in units of $k_B$/unit-cell. 
The logarithmic specific heat singularity of the Ising model ($K_4 = 0$), is eliminated and replaced by 
a bump whose width increases as  $|K_4|$ increases. The arrow in the lower right panel indicates $T_c$ 
as obtained from the peak in the susceptibility $\chi_s$.}
\label{CV_fig1}
\end{figure}

 
We first consider the case of isotropic Ising coupling $K_l=K_t$, i.e. $A=1.0$, next-nearest neighbor 
coupling $K^{xy}=0.0$, and increasing $|K_4|$. We use this case for reference, as this parameter set 
represents the standard isotropic AT model \cite{AT,Aji-Varma}. The results for the specific heat are 
shown in Fig. \ref{CV_fig1}. The logarithmic specific heat of the Ising model disappears to be replaced 
by a bump whose extent in $T$ increases as  $|K_4|$ increases. This is consistent with the asymptotic 
critical exponents \cite{AT}. 
\par
\begin{figure}[htbp]
\centerline{\hbox{\includegraphics[width=80mm]{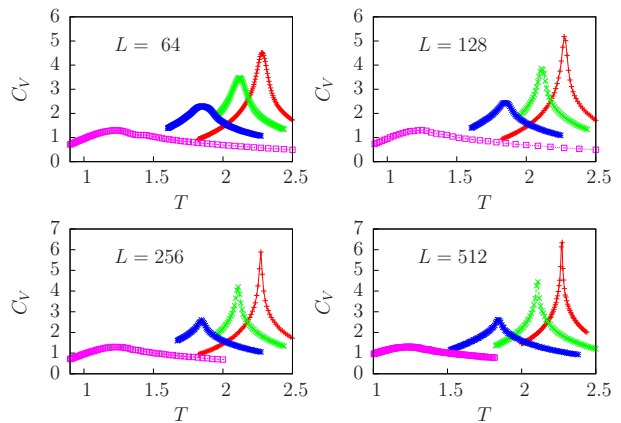}}}
\caption{(Color online) Specific heat as a function of temperature $T$ for the classical part 
of the generalized AT model Eq. \ref{Model}, with $A=1.0$ and $K^{xy}=0.0$, for various values of 
$K_4 = 0.0,-0.1,-0.25,-0.5$, and system size $L=64,128,256,512$. The vertical scale is in units of 
$k_B$/unit-cell. The logarithmic specific heat singularity of the Ising model ($K_4 = 0$), is 
eliminated and replaced by a bump whose width increases as  $|K_4|$ increases. Note how the
double-bump in $C_V$, which is present at smaller system sizes, disappears when $L$
is increased. When $L=512$, the results appear to be well converged.}
\label{multi_cv}
\end{figure}

In Fig. \ref{multi_cv}, we investigate how well these results are converged when increasing
the system size through the values $L=64,128,256,512$. It is seen that the results appear well
converged when $L$ has reached $256$, in particular the double-peak structure in $C_V$ that 
is present for small system sizes disappears upon increasing $L$. In contrast to the 
Binder-cumulant (see below), we have not attempted a data collapse of the specific heat by 
trying a scaling form $C_V(T,L) = L^{\alpha/\nu} {\cal C}_{\pm} (L^{1/\nu}(T-T_c)/T_c)$
and adjusting $\alpha$ to obtain data-collapse. The reason is that we anticipate a negative 
specific heat exponent, such that corrections to the above scaling form will be large, thus 
preventing data collapse. Even for positive $\alpha$, it is well-known that corrections to
scaling are substantial for the specific heat. This simply means that the specific heat 
by itself oddly enough is not a very useful quantity from which to extract precise values of 
$\alpha$ in Monte-Carlo computations on practical system sizes. Other techniques are required for 
this, see e.g. Ref. \onlinecite{third_moment}. However, the main point of the present paper is not 
to determine a precise value of $\alpha$ numerically, but rather to demonstrate (including all 
corrections to scaling) that a striking suppression of the prominent logarithmic singularity 
of the $2D$ Ising model takes place as $|K_4|$ is increased. Fig. \ref{multi_cv} clearly shows 
that the suppression is not a finite-size artifact. Note in particular that the relative height 
of the bump in $C_V$ for non-zero $|K_4|$ is suppressed compared to the Ising-singularity as 
$L$ increases. 
\par
\par
\begin{figure}[htbp]
\centerline{\hbox{\includegraphics[width=80mm]{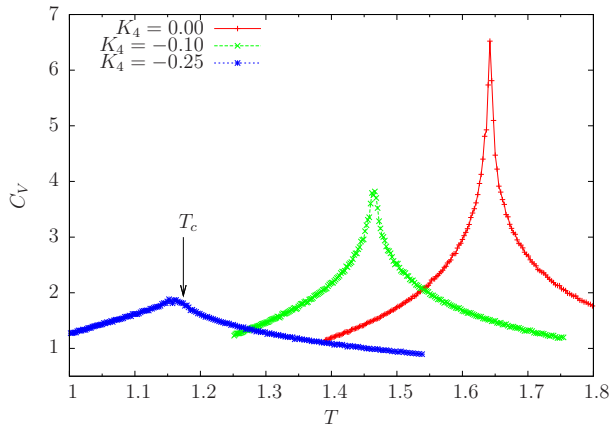}}}
\caption{(Color online) Specific heat  as a function of  temperature $T$ for 
the classical part of the generalized AT model Eq. \ref{Model}, with $A=0.5$ and $K^{xy}=0.0$, 
for various  values of $K_4 = 0.0,-0.1,-0.25$, and system size $L=512$. The vertical 
scale is in units of $k_B$/unit-cell. Compared to the case shown in Fig. \ref{CV_fig1}, with 
$A=1.0$, precisely the same trends are seen in the evolution of the anomaly as the AT coupling 
$|K_4|$ is increased, only slightly more pronounced. The arrow  indicates $T_c$ as obtained 
from the peak in the susceptibility $\chi_s$.}
\label{CV_fig2}
\end{figure}

We next consider the effect of increasing the anisotropy  ($A < 1$), such as to weaken the 
ordering in each of the $J_{y}({\bf r})$- and $J_{x}({\bf r})$ Ising fields. Note, however, 
that because the anisotropy introduced is equal for both of the Ising fields (only the 
direction of the anisotropy is changed) the model only has one single critical point even 
in the absence of a $K_4$-coupling. The model is then merely two copies of one and the same 
anisotropic $2D$ Ising model. However, an increase in $|K_4|$ is expected to have a stronger 
effect for $A < 1.0$ than when $A=1.0$ due to the weaker ordering and reduced critical 
temperature. The bump in the specific heat is then accordingly smoother as seen in Fig. 
\ref{CV_fig2} compared to Fig. \ref{CV_fig1}.    
\par
\begin{figure}[htbp]
\centerline{\hbox{\includegraphics[width=80mm]{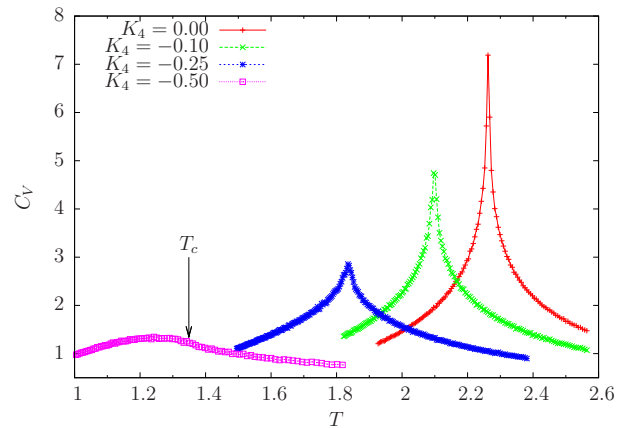}}}
\caption{(Color online) Specific heat  as a function of temperature $T$ for the 
classical part of the generalized AT model Eq. (\ref{Model}), with $A=1.0$ and 
$K^{xy}=0.1$, for various  values of $K_4 = 0.0,-0.1,-0.25,-0.5$, and system size 
$L=512$. The arrow indicates $T_c$ as obtained from the peak in the susceptibility 
$\chi_s$. The vertical scale is in units of $k_B$/unit-cell.}
\label{CV_fig3a}
\end{figure}
\begin{figure}[htbp]
\centerline{\hbox{\includegraphics[width=80mm]{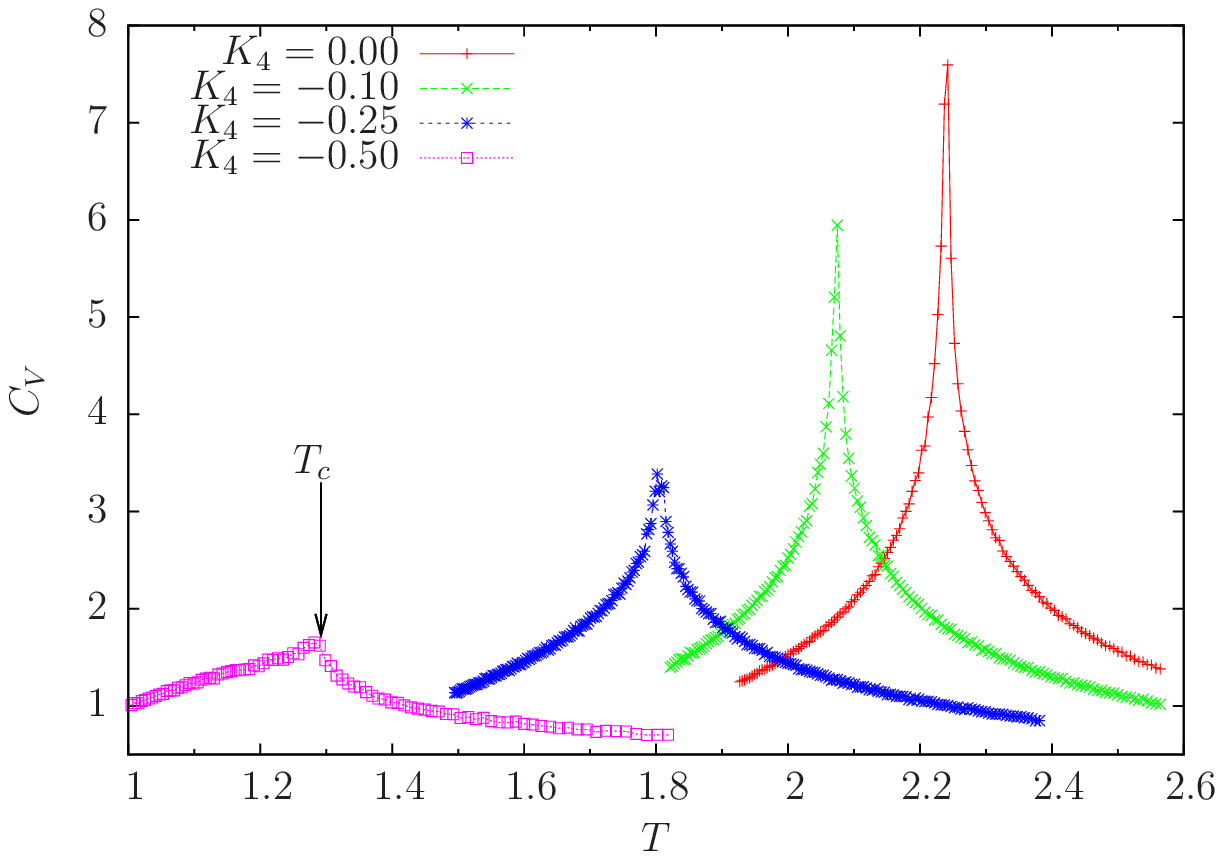}}}
\caption{(Color online) Specific heat  as a function of temperature 
$T$ for the classical part of the generalized AT model Eq. \ref{Model}, 
with $A=1.0$ and $K^{xy}=0.2$, for various  values of $K_4 = 0.0,-0.1,-0.25,-0.5$, 
and system size $L=512$. The arrow indicates $T_c$ as obtained from the peak 
in the susceptibility $\chi_s$. The vertical scale is in units of $k_B$/unit-cell.}
\label{CV_fig3b}
\end{figure}
\begin{figure}[htbp]
\centerline{\hbox{\includegraphics[width=80mm]{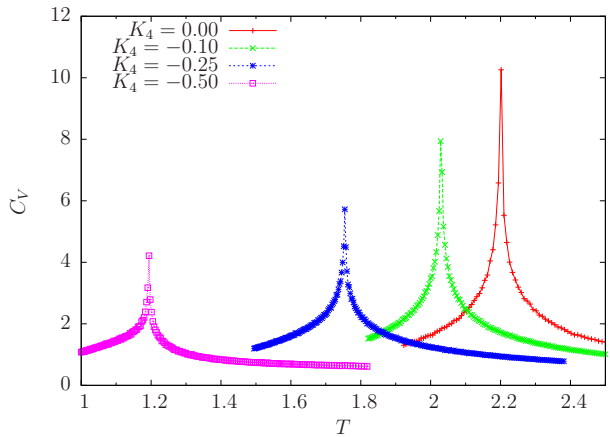}}}
\caption{(Color online) Specific heat anomaly as a function of temperature 
$T$ for the classical part of the generalized AT model Eqs. (\ref{Model}), 
with $A=1.0$ and $K^{xy}=0.3$, for various  values of $K_4 = 0.0,-0.1,-0.25,-0.5$, 
and system size $L=512$. The vertical scale is in units of $k_B$/unit-cell.}
\label{CV_fig3c}
\end{figure}

We now repeat the above computations for $A=1.0$ with $K^{xy}=0.1,0.2$ and $0.3$. This coupling 
tends to frustrate the Ising ordering, since a large $K^{xy}$ tends to promote striped order  due 
to the diagonal anisotropy (represented by a change of sign in $K^{xy}$ upon $\pi/2$ rotations of 
next-nearest neighbor vectors). It is of interest to see how the presence of $K^{xy}$ affects the 
introduction of  the AT coupling $K_4$. Naively, since the coupling $K^{xy}$ promotes striped order 
and frustrates the uniform order promoted by $K_x,K_y$, we would expect that the suppressed anomalies 
are pushed to lower temperatures as $K^{xy}$ is increased. In Figs. \ref{CV_fig3a}, \ref{CV_fig3b}, 
and \ref{CV_fig3c}, we show the  specific heat  for the same sets of parameters as in Fig. \ref{CV_fig1}, 
except that now $K^{xy}=0.1,0.2,0.3$, respectively. 

We see that the effect of $K^{xy}$ is to increase the sharpness of the bump in the 
specific heat, while the effect of $K_4$ again is to widen the bump (in the presence 
of $K^{xy}$). We also see that the anomalies that remain are pushed slightly downwards 
in temperature compared to the case $K^{xy}=0$, cf. the results of  Fig. \ref{CV_fig1}. 
The change is however only minor for the cases $K^{xy}=0.1$ and $K^{xy}=0.2$, consistent 
with the weak suppression of the critical temperature we found upon increasing $K^{xy}$ 
at $K_4=0$ in Fig. \ref{CV_fig0}. The conclusion we draw from these computations is 
that  the singularity of the specific heat of the Ising case is removed by the coupling 
$K_4$ is included. The resulting bump in the specific heat becomes sharper for increasing 
$K^{xy}$ at finite $K_4$.  
\par
\begin{figure}[htbp]
\centerline{\hbox{\includegraphics[width=80mm]{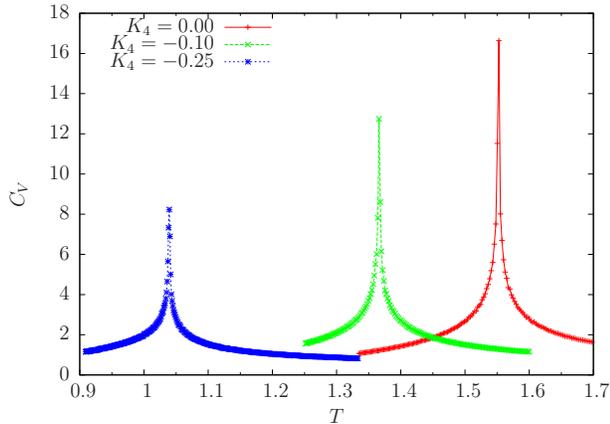}}}
\caption{(Color online) Specific heat anomaly as a function of  temperature 
$T$ for the classical part of the generalized AT model Eq. \ref{Model}, with 
$A=0.5$ and $K^{xy}=0.3$, for various  values of $K_4 = 0.0,-0.1,-0.25$, and 
 system size $L=512$. The vertical scale is in units of $k_B$/unit-cell.} 
\label{CV_fig4}
\end{figure}

Finally, we consider the most general case of anisotropic Ising coupling $A =0.5$ and 
finite $K^{xy}=0.3$, as $|K_4|$ is increased, shown in Fig. \ref{CV_fig4}.  
It is clear from Fig. \ref{CV_fig4} that the introduction of anisotropy $A = K_t/K_l =0.5$
widens the width of the bump in the specific heat. This is easily understood, since 
increasing anisotropy implies that the magnitude of $K_4$ relative to the Ising 
couplings in the problem will increase. The effect of a given increase in $K_4$ is 
therefore  more strongly felt. Moreover, as in the isotropic case, the anomalies are 
pushed down in temperature compared to the case $K^{xy}=0$, cf. the results of 
Fig. \ref{CV_fig2}.
\par
Concluding this section on the results for the specific heat, we mention that we have also,
at the early stages of this work, performed a rather rudimentary comparative study of the 
specific heat anomaly in the $2D$ Ashkin-Teller model and the $2DXY$ continuous rotor model 
with a $4$-fold symmetry breaking term, on lattice sites up to $L=32$. This numerics is 
insufficient to draw any conclusions about the fluctuation spectrum on the disordered side 
of the transition, close to the transition, as the symmetry breaking field becomes small. That 
is, the simulations {\it per se} do not allow us to conclude anything about the perturbative 
relevance or irrelevance of the symmetry breaking term. What we have been able to confirm, is 
that the specific heat anomaly of the $2D$ Ashkin-Teller model is indistinguishable from the 
$2DXY$ continuous rotor model with a symmetry breaking term, provided the symmetry breaking 
term is large. 

\subsection{$\Mso$, $\chimso$, and the critical exponents $\nu$ and $\beta$}\label{ms_chi}
\begin{figure}[htbp]
\centerline{\hbox{\rotatebox{0}{\includegraphics[width=85mm]{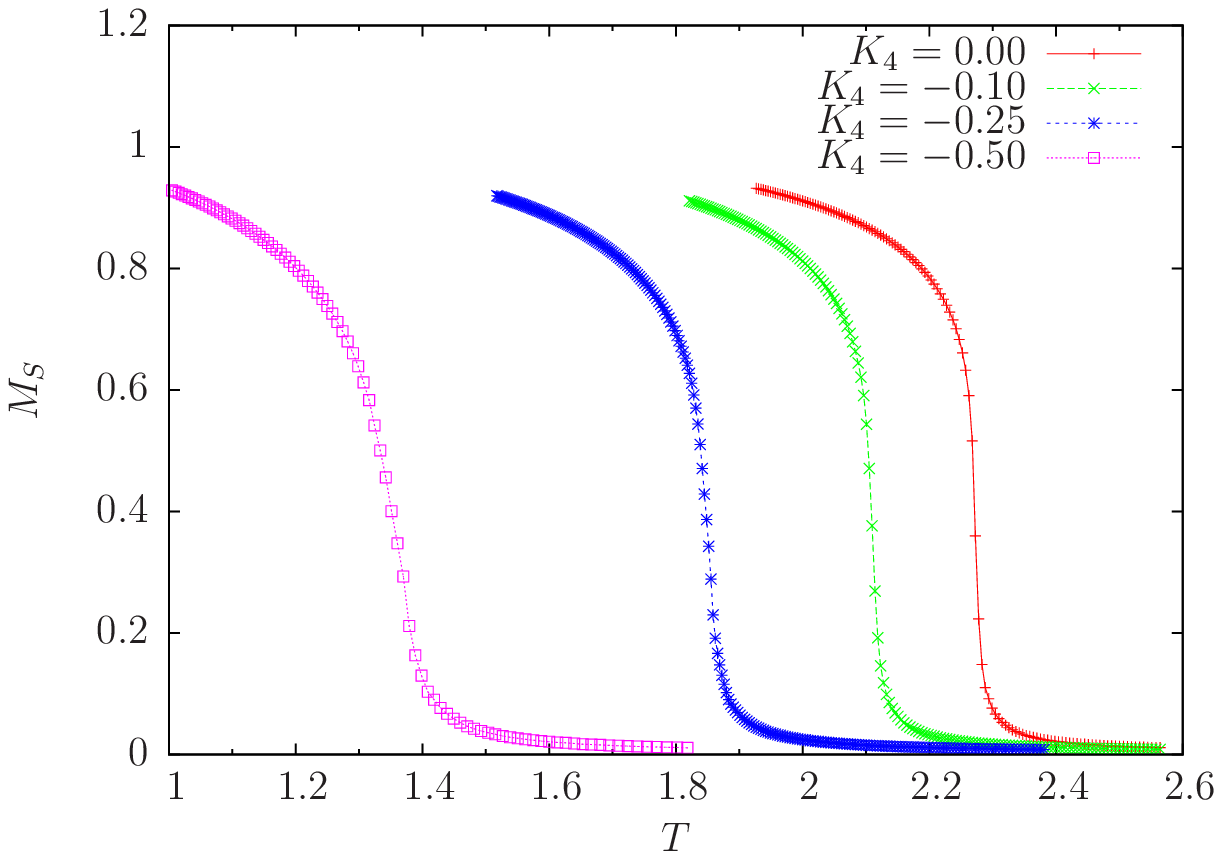}}}}
\caption{(Color online) The staggered order parameter, Eq. \ref{stag_mag}, as a function of 
temperature $T$ for the classical part of the generalized AT model Eq. \ref{Model}, 
with $A=1.0$ and $K^{xy}=0.0$, for various  values of $K_4 = 0.0,-0.1,-0.25,-0.5$, and system 
size $L=512$.}
\label{stagmag_fig1}
\end{figure}

\begin{figure}[htbp]
\centerline{\hbox{\rotatebox{0}{\includegraphics[width=85mm]{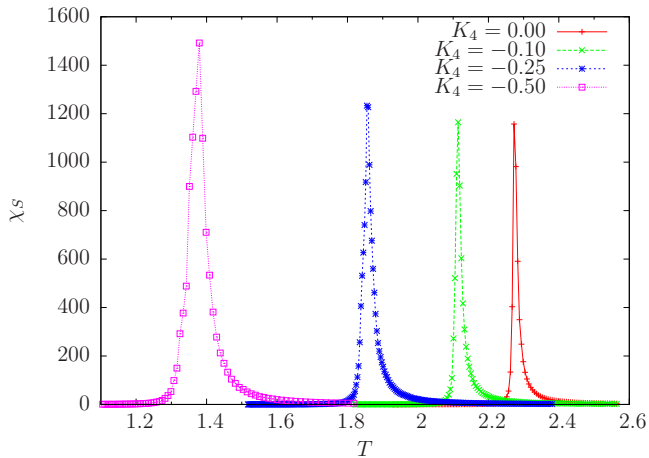}}}}
\caption{(Color online) The susceptibility of the staggered magnetization within each unit cell, 
Eq. \ref{chi_mso}, as a function of temperature $T$  for the classical part of the generalized 
AT model Eq. \ref{Model}, with $A=1.0$ and $K^{xy}=0.0$, for various values of $K_4 = 0.0,-0.1,-0.25,-0.5$, 
and system size $L=512$. Note that the susceptibility retains the non-analytical features of the 
Ising-case even for parameters where the specific heat anomaly is completely suppressed.}
\label{chim_fig1}
\end{figure}

\begin{figure}[htbp]
\centerline{\hbox{\rotatebox{0}{\includegraphics[width=85mm]{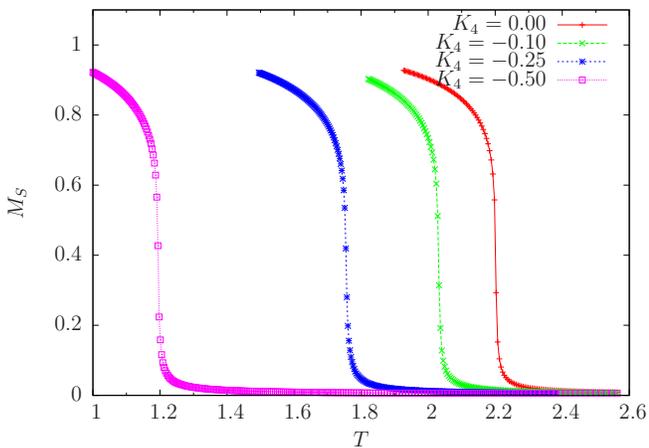}}}}
\caption{(Color online) The staggered order parameter, Eq. \ref{stag_mag}, as a function of 
temperature $T$ for the classical part of the generalized AT model Eq. \ref{Model}, with $A=1.0$ 
and $K^{xy}=0.3$, for various  values of $K_4 = 0.0,-0.1, -0.25,-0.5$, and system size $L=512$.}
\label{stagmag_fig2}
\end{figure}

\begin{figure}[htbp]
\centerline{\hbox{\rotatebox{0}{\includegraphics[width=85mm]{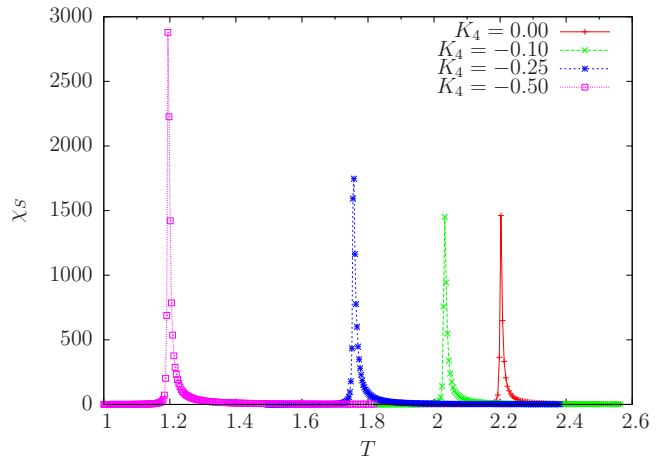}}}}
\caption{(Color online) The susceptibility of the staggered magnetization within each 
unit cell, Eq. \ref{chi_mso}, as a function of temperature $T$  for the classical part 
of the generalized AT model Eq. \ref{Model}, with $A=1.0$ and $K^{xy}=0.3$, for various 
values of $K_4 = 0.0,-0.1,-0.25,-0.5$, and system size $L=512$. Note the marked increase
in the susceptibility as $-K_4$ is increased, in contrast to the suppression of the
anomaly in the specific heat.}
\label{chim_fig2}
\end{figure}

Let us now study the order parameter and susceptibility of the order parameter, $\Mso$ 
and $\chimso$, Eqs. \ref{stag_mag} and \ref{chi_mso}. We have first chosen parameters $A=1.0$, 
$K^{xy}=0$, and varied $K_4$, for which the evolution of the specific heat anomaly is shown in 
Fig. \ref{CV_fig1}. The results for $\Mso$ and $\chimso$  are shown in Figs. 
\ref{stagmag_fig1} and \ref{chim_fig1}, respectively. 
We see that the staggered magnetization retains a non-analytic behavior as in the pure Ising 
case even for $K_4=-0.5$. This contrasts sharply with the lack of any traces of singular
behavior in the specific heat, cf. Fig. \ref{CV_fig1}. From Fig. \ref{chim_fig1} we see the 
same trend, namely that the susceptibility retains a non-analytic feature even for the largest 
$K_4$ values we have considered, and which suffice to completely suppress the singularity in 
the specific heat. 
We have repeated these calculations with $K^{xy}=0.3$. The results are shown in Figs. \ref{stagmag_fig2} 
and \ref{chim_fig2}, with essentially the same results as in Figs. \ref{stagmag_fig1} and \ref{chim_fig1}.   

\par
\begin{figure}[htbp]
\centerline{\hbox{\includegraphics[width=80mm]{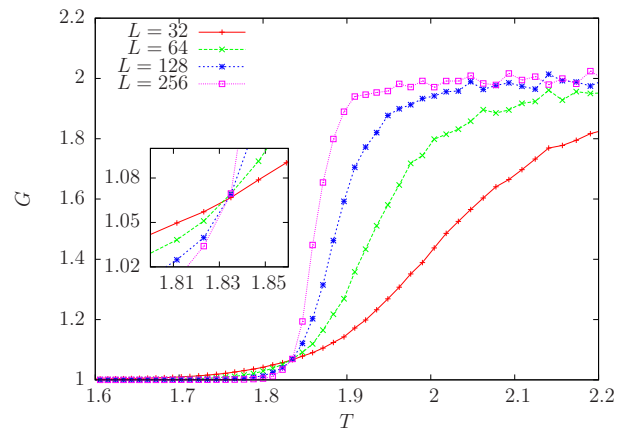}}}
\caption{(Color online) The Binder-cumulant $G$, Eq. \ref{Binder}, as a function
of $T$ for the model 
Eq. \ref{Model}, for $A=1.0,K^{xy}=0.1,K_4=-0.25$ for various system sizes,
in the absence of Ferrenberg-Swendsen reweighting. The inset 
shows a blowup of the temperature-region where the lines for various system sizes 
cross, providing an estimate for $T_c$.}
\label{Binder_plain}
\end{figure}
\begin{figure}[htbp]
\centerline{\hbox{\includegraphics[width=80mm]{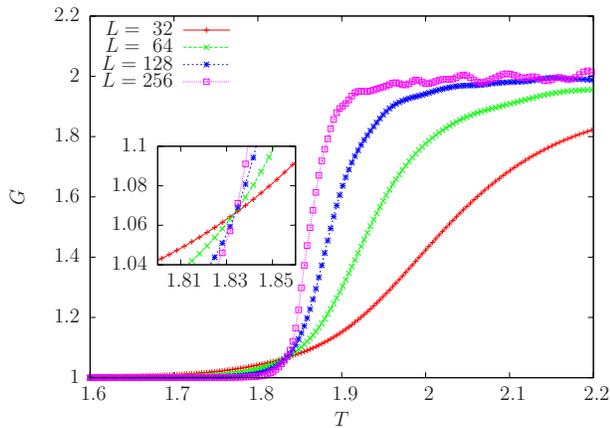}}}
\caption{(Color online) The Binder-cumulant $G$, Eq. \ref{Binder}, as a function
of $T$ for the model Eq. \ref{Model}, for $A=1.0,K^{xy}=0.1,K_4=-0.25$ for various 
system sizes, using Ferrenberg-Swendsen reweighting. The inset shows a blowup of 
the temperature-region where the lines for various system sizes cross, providing 
an estimate for $T_c$. Note the consistency of the  estimate for $T_c$ compared to 
Fig. \ref{Binder_plain}.}
\label{Binder_RW}
\end{figure}

We next attempt to estimate the critical exponents $\nu$ and $\beta$ for the model 
Eq. \ref{Model}, for the set of parameters $A=1.0,K^{xy}=0.1,K_4=-0.25$. (For the same 
set of parameters, but $K^{xy}=0$ and $K_4=0$, see comments below). We base our 
calculations of these critical exponents on using the Binder cumulant Eq. \ref{Binder} and 
the scaled staggered magnetization $L^{\beta/\nu} \Mso$, cf. Eq. \ref{stag_mag}. For these 
computations, we have used up to $3 \cdot 10^6$ sweeps over the lattice for each temperature.
In addition, we have used Ferrenberg-Swendsen (FS) multihistogram reweighting \cite{FSH} of the 
raw data for the Binder cumulant in order to improve on the accuracy. The method of computation 
is described in Section \ref{crit_exp}. In Fig. \ref{Binder_plain}, we show the Binder cumulant 
for various system sizes as a function of the temperature $T$ without reweighting. The crossing 
points provide an estimate for $T_c$. Even in the absence of FS reweighting, there is very
little scatter in 
these crossing points, and $T_c$ is determined with an uncertainty of much less than $1 \%$. With 
Ferrenberg-Swendsen reweighting, this picture remains, as is seen from Fig. \ref{Binder_RW} 
where reweighting is used. An accurate estimate for $T_c$ will turn out to be crucial in the 
following. Note also that the estimates we get for $T_c$ from the crossing of lines in the Binder 
cumulant are well in agreement from the somewhat cruder estimates we would obtain from determining 
the temperatures  at which the peaks of the staggered susceptibilities occur. 

\par
\begin{figure}[htbp]
\centerline{\hbox{\includegraphics[width=80mm]{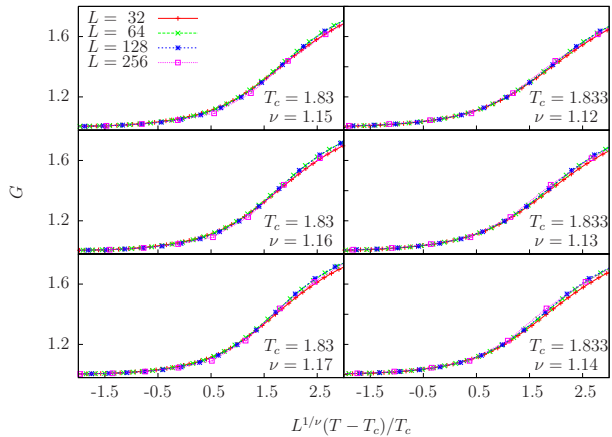}}}
\caption{(Color online) The Binder-cumulant $G$, Eq. \ref{Binder} as a function
of the quantity $L^{1/\nu} (T-T_c)/T_c$, for the model Eq. \ref{Model}, for 
$A=1.0,K^{xy}=0.1,K_4=-0.25$ for various system sizes. Ferrenberg-Swendsen reweighting 
of the data is used.  We have taken estimates for $T_c$ from Fig. \ref{Binder_RW} and 
adjusted the correlation length critical exponent $\nu$ to achieve the best data collapse.
As is seen, the optimal $\nu$ is extremely sensitive to the chosen value of $T_c$.}
\label{Scaled_Binder_Kxy_01}
\end{figure}

In Fig. \ref{Scaled_Binder_Kxy_01}, we replot the same Binder-cumulant, now as 
a function of the  quantity $L^{1/\nu}(T-T_c)/T_c$, using estimates for $T_c$ from Fig. \ref{Binder_RW}
and adjusting $\nu$ to get data collapse. While we see that the above computations do 
not allow us to extract extremely precise values of $\nu$, it does allow us to conclude 
that the exponent $\nu$ is consistent with the values obtained from the Ashkin-Teller 
model, and that $\nu$ appears to be enhanced compared to the $2D$ Ising value $\nu = 1$.  
\par
\begin{figure}[htbp]
\centerline{\hbox{\includegraphics[width=80mm]{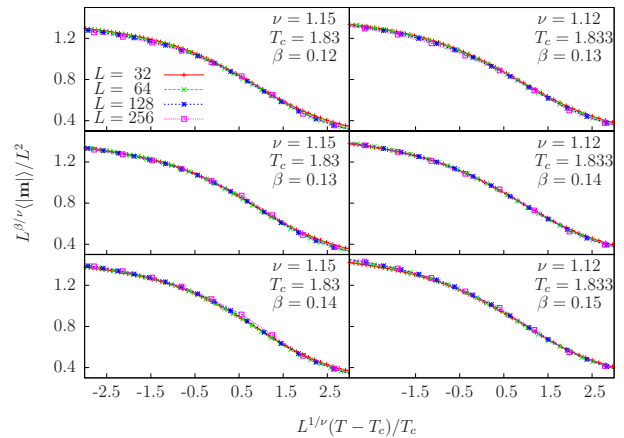}}}
\caption{(Color online) The scaled staggered order parameter $L^{\beta/\nu} \Mso$, 
cf. Eq. \ref{stag_mag}, as a function of the quantity $L^{1/\nu} (T-T_c)/T_c$, for the 
model Eq. \ref{Model}, for $A=1.0,K^{xy}=0.1,K_4=-0.25$ for various system sizes. 
Ferrenberg-Swendsen reweighting of the data is used. We have  taken estimates for $T_c$ 
from Fig. \ref{Binder_RW} and estimates for $\nu$ from Fig. \ref{Scaled_Binder_Kxy_01} 
and adjusted the order-parameter exponent $\beta$ to achieve the best data collapse. }
\label{Scaled_StagMag_Kxy_01}
\end{figure}

We next compute the quantity $L^{\beta/\nu} \Mso$ as a function of the quantity 
$L^{1/\nu} (T-T_c)/T_c$ to obtain the order parameter exponent $\beta$, by using the 
values of $T_c$ and $\nu$ obtained from the scaled Binder cumulant in Fig. \ref{Scaled_Binder_Kxy_01}, 
and then adjusting $\beta$ to get data collapse of all magnetization curves for various 
values of $L$. The result of this procedure is shown in Fig. \ref{Scaled_StagMag_Kxy_01}. 
Again, from the above we cannot conclude anything with great precision about the exponent 
$\beta$, other than saying that it is consistent with the exact values that are known for 
the Ashkin-Teller model, i.e. Eq. \ref{Model} with $K^{xy}=0$.

We have also checked the exponents for the same set of parameters as above, except that 
$K^{xy}=0$. We draw the conclusion that to the level of precision of the above computations, 
the exponents are not altered from the Ashkin-Teller case. However, when we repeat the 
procedure for the same set of parameters as above, except that $K^{xy}=0.3$, we find 
that there is a clear deviation and that the exponents $\nu$ and $\beta$ definitely  do 
not take Ashkin-Teller values. In particular, we get optimum data collapse for $\beta$ 
clearly less then $1/8$. From this, we infer that while the parameter $K^{xy}$ may be 
perturbatively irrelevant, it may alter the universality class of the phase transition 
of the model if it is large enough. We also note that the reason that $K^{xy}$ appears to 
have much less of an effect on the transition when $K_4=0$ compared to when $K_4=-0.25$, is 
that the latter case represents a frustration of the ferromagnetic Ising ordering that 
lowers the critical temperature of the system and  enhances the effect of introducing 
$K^{xy}$, which also frustrates the ferromagnetic Ising ordering, and  promotes striped 
ordering.
\par
Non-universality in $\beta$ due to the presence of the parameter $K_4$ in the problem means 
that $\beta$ in principle should vary slightly as we cross the pseudogap line vertically in 
the $(x,T)$-phase diagram of high-$T_c$ cuprates as the doping is varied, if we assume that 
the parameters of the effective model Eq. \ref{Model} varies as we move along the pseudogap 
line. In particular, a variation of $\beta$ with $K_4$ is clearly seen from Fig. \ref{stagmag_fig1},
although we have not performed a detailed finite-size scaling analysis to determine $\beta$ 
as a function of $K_4$. We also note from Fig. \ref{CV_fig0} that introduction of $K^{xy}$ 
does not change the universality class of the transition when $K_4 = 0$. We may therefore 
quite reasonably assume that the presence of $K^{xy}$ does not change the Ashkin-Teller 
universality class of the phase transition when $K_4$ is present, provided $K^{xy}$ is not 
too large. We may then deduce that for negative $K_4$, we will have $-2 < \alpha < 0$, $1/8 < \beta < 1/4$, 
and $7/4 < \gamma < 7/2$. A suppression of the specific heat anomaly as seen for the case $K_4=-0.25$, 
puts us at $\alpha \approx -0.37$, $\beta \approx 0.15$, and $\gamma \approx  2.07$. The weak 
variation in the exponent $\beta$ from the Ising value $1/8$ is due to the near-cancellation of 
the rather large, but opposite, variations in the specific-heat 
exponent $\alpha$ and the susceptibility-exponent $\gamma$, consistent with the scaling law 
$\alpha + 2 \beta + \gamma = 2$. It is precisely the large variation in $\alpha$ that wipes out 
the specific-heat anomaly that also produces a large enhancement of the susceptibility of the 
staggered orbital magnetization, see Fig. \ref{exponents}.

\subsection{Comparison of Calculated Specific Heat with Experiments} 

We use the results in Figs. \ref{CV_fig0}-\ref{CV_fig4} to estimate the peak value of the specific heat bump 
expected due to the transition to see why it may be unobservable in experiments performed so far. In comparing 
with experiments, the following should be borne in mind.  The ordering below the transition temperature is 
three-dimensional. Hence, the observed specific heat will be that of the form calculated above as temperature 
is decreased towards $T^*$, followed by a singularity characteristic of the $3D$ Ising model near $T^*$ and 
below it. However, the integrated value of specific heat divided by $T$ under the singularity is only a fraction 
of the total entropy due to the loop order degrees of freedom. As we discuss below, the latter itself is more 
than an order of magnitude smaller than the entropy due to fermionic excitations in the same temperature range. 

The area $\int (C_v(T)/T) dT$ over all temperatures in each of the curves in Figs. \ref{CV_fig0}-\ref{CV_fig4}
is $2 \ln (2)k_B$/unit-cell, reflecting that the calculations are performed for 2 Ising degrees per unit-cell. 
Given that the ordered moment due to orbital currents is estimated in neutron scattering experiments to be 
$10^{-1} \mu_B$/unit-cell, the integrated value is expected to be $2 \ln (2)k_B$/unit-cell multiplied by $O(10^{-2})$. 
To compare with experiments, we may consider calculations for the case $K^{xy} =0$ and $|K_4/K_l|$ between, say  $0.25$ 
and $0.5$. The peak value of the specific heat from Figs. \ref{CV_fig0}-\ref{CV_fig4} is then expected to be less 
than  $0.5 \times 10^{-2} k_B$ per unit-cell or less than about $0.05$ Joules/mole/degree. This should be compared 
with the measured specific heat \cite{Loram}, which at about 200 K is about 200 Joules/mole/degree. In Ref. 
\onlinecite{Loram}, the electronic specific heat is estimated by subtracting the specific heat for a similar 
non-metallic compound to be about $2$ Joules/mole/degree. Therefore, the bump has a peak which is  $3$ to $4$ 
orders of magnitude smaller than the total specfic heat, and $1$ to $2$ orders of  magnitude smaller than even 
the deduced electronic specific heat. Given that the specific heat bump is spread out over temperatures of  
${\cal{O}}(2 T^*)$, it is not surprising that with pseudogap temperatures of ${\cal{O}}(200)$ K or higher, it 
has gone undetected. There are underdoped cuprates with lower $T^*$, in which a bump in the specific heat with 
magnitude of order that suggested here is claimed \cite{Monomo} to be observed. 

\section{Uniform Susceptibility}\label{uniform-chi}
Just as the onset of antiferromagnetic spin-order has a weak parasitic non-analytic effect on 
the uniform magnetic susceptibility, the onset of loop-current orbital magnetic order may be 
expected to have a similar effect on the uniform magnetic susceptibility. Such an effect has 
indeed been measured recently in careful studies across $T^*(x)$ \cite{Leridon}.  
\par
Since the uniform magnetization is a parasitic effect on the staggered magnetization, we can 
calculate its temperature dependence by a Landau theory in which we consider the free energy 
for the staggered magnetization, but consider the minimal coupling of the uniform magnetization 
to the staggered magnetization.  Let $\Mso$ be the staggered magnetization and $\langle M \rangle$ 
be the thermal average of the uniform magnetization in the presence of an external field $H$. Let 
$F_0(M_{s0})$ be the free-energy for the $\Mso$ in the absence of an external magnetic field $H$. 
Quite generally, the leading terms in the free-energy are given by
\be
\label{FE}
F= F_0(M_s) + \frac{M^2}{2\chi_0} -M H + \frac{C}{2} M_s^2 M^2 +...
\ee
Here, $C$ is a coefficient which gives the competition between the staggered magnetization and uniform 
magnetization. The sign of $C$ is positive if as is reasonable, the  staggered magnetization  decreases 
if the uniform magnetization increases, and vice versa. 

\par
This form of the free energy gives correct answers only in the regime in which the staggered susceptibility 
is small and therefore is not valid very close to the transition. Also, the susceptibility calculated is 
for magnetic field parallel to the direction of sub-lattice magnetization. In the simplest theory, this 
direction is perpendicular to the Cu-O planes. In the experiments \cite{Fauque}, an angle closer to 
$\pi/4$ has been deduced for which some theoretical justifications are provided \cite{Aji-Varma2, Weber}. 
Since the experiments are done in powder samples, we will ignore this issue for the present. 

\par 
Let $\chi_{s0} \equiv (\partial^2F_0/\partial M_{s0}^2)^{-1}$ be the order parameter susceptibility, 
which is calculated above. The subscript $0$ in $\chi_{s0}$ indicates the quantity in the absence of 
$\langle M \rangle$.  $\chi_0$ is the uniform susceptibility in the absence of $M_{s}$. Then in the 
presence of $\langle M \rangle$, induced by the external field $H$, the condition
\be
    \frac{\partial F}{\partial M} =0 ,
\ee
gives
 \be
\frac{ \langle M\rangle }{\chi_0} - H + C \langle M\rangle  \langle \Mso^2\rangle  = 0.
 \ee
This gives $\langle M\rangle  \equiv \chi H$ in linear response (i.e. low $H$), with 
 \be
 \chi = \frac{\chi_0}{1+C\chi_0\langle \Mso^2\rangle }.
 \label{chi-uni}
 \ee
Here, $\langle M_s^2 \rangle$ is the thermodynamic squared magnetization in the presence 
of $\langle M \rangle $, and $\chi$ is the uniform susceptibility.  We may write Eq. 
\ref{chi-uni}
as 
\begin{eqnarray}
\chi = \frac{\chi_0}{1+ C \chi_0(\langle M_s \rangle^2 + T \chi_s) }.
\label{chi-uni2}
\end{eqnarray}
Also, quite generally, the order parameter susceptibility is
 \be
 \label{chis1}
 \chi_s^{-1} = \frac{\partial^2F}{\partial M_s^2} =  \chi_{s0}^{-1} + C <M^2> = \chi_{s0}^{-1}  + CT\chi.
 \ee
 This  gives 
 \be
 \label{chis2}
 \chi_s = \frac{ \chi_{s0}}{1+CT\chi \chi_{s0}}. 
 \ee
Thus, using Eqs.~(\ref{chis1},\ref{chis2}), we may write $\chi$ in terms of known quantities 
$\chi_0$ and $\chi_{0s}$, to obtain
 \be
 CT\chi_{s0}\chi^2 + \chi -\chi_0 =0.
 \ee
For $T>>T_c$, where the above treatment is valid, we have  $4CT\chi_{s0}\chi_0 << 1$, so that
  \be
  \label{chi3}
 \chi \approx \chi_0 - CT \chi_{s0}\chi_0^2, ~~ (T-T_c)/T_c >>1.
 \ee
The uniform susceptibility is therefore predicted to decline from its constant Pauli value at far above $T_c$ in 
the same range that $\chi_{s0}$ shows a rise. We suggest that the observed slow decrease of $\chi(T)$ for 
temperatures well above $T^*$ be fitted to such a form.
\par
Well below $T_c$, the model behaves as an Ising model. Therefore, the contribution of the ordered moments to 
the uniform  susceptibility approaches zero exponentially as $ T \to 0$.

\subsection{Mean-field Jump in $d\chi/dT$ at $T_c$}\label{chis} 
In a mean-field calculation $\chi_{s0}$ does not change above $T_c$. There is, however, a jump in $d\chi/dT$ 
expected at $T_c$. The experimental results  have been quantified by such a jump \cite{Leridon}. To compare 
with available experimental results,  we approximate Eq.~(\ref{chis1}) as
\be
 \label{chi4}
 \chi \approx \chi_0 - C\chi_0^2<M_s>^2,
 \ee
 so that
 \be  
 \label{d/dTchi}
 \frac{d\chi}{dT} = -C\chi_0^2 \frac{d<M_s>^2}{dT}.
 \ee
Here, a temperature independent $\chi_0$ is assumed. We now need to know the right side of Eq.~(\ref{d/dTchi}). 
This may be estimated as follows.  Returning to Eq.~(\ref{FE}), we may write $F_0(M_s)$ as
\be
\label{F0}
F_0(M_s)= \tilde{\alpha}/2\frac{(T-T_0^*)}{T_0^*}M_s^2 + \frac{\beta}{4} ~ M_s^4 +....
\ee
This defines the transition temperature $T_0^*$ in the absence of an external magnetic field $H$, i.e. for $M=0$. It also defines an inverse 
susceptibility $\alpha$ for $M_s$, which we expect to be of the same order as to the inverse of the density of states at the 
Fermi-surface, or equivalently of order $\chi_0^{-1}$. Combining Eq.~(\ref{F0}) with the third term in Eq.~(\ref{FE}), we see 
that a finite $M$ leads to a decrease in the transition temperature $\delta T^*$, with
 \be
 \label{dT}
 \frac{\delta T^*}{T^*} \approx C M^2/\tilde{\alpha}.
 \ee
 Note also that 
 \be
 M_s^2 \approx M_{s}(0)^2\frac{(T_0^*-T)}{T_0^*},
 \ee
 where $M_{s}(0)^2$ is the zero temperature value of $M_s^2$.
 Using this in Eq.~(\ref{d/dTchi}), the jump in the derivative of the susceptibility at $T^*$ is given by
 \be
 \frac{T_0^*}{\chi_0}\frac{d \chi}{dT} = CM_s(0)^2 \chi_0.
  \ee
Now we need an estimate of $C M_s(0)^2$. This can be obtained from Eq.~(\ref{dT}) if we note that the 
transition temperature will be reduced to 0, i.e., $\frac{\delta T^*}{T^*} =1$, for some magnetization 
$M*$. The magnitude of $M*$ has to be the same order as $M_s(0)$ at zero field. Therefore
 \be
 \label{C}
 C M_s(0)^2/\tilde{\alpha} \approx 1.
 \ee
Using this above, the jump in $d\chi/dT$ at $T^*$ is given by
 \be
 \label{final}
 \frac{T_0^*}{\chi_0}\frac{d \chi}{dT} \approx \tilde{\alpha} \chi_0 \approx 1,
 \ee
where we have used the estimate for $\tilde{\alpha}$ estimated earlier.
\par 
In the experiments reported in Ref. \onlinecite{Leridon}, a value of  $\frac{T_0^*}{\chi_0}\frac{d \chi}{dT}$ between 
$0.2$ and $0.3$ has been deduced. This should be considered in good agreement with the estimate of $O(1)$. The weak 
assumptions in the analysis above are the lack of knowledge of the numerical constant between $\tilde{\alpha}$ and 
$\chi_0^{-1}$, and the unknown numerical constant on the right hand side of Eq.~($\ref{C})$, instead of $1$. However, 
they cannot be off by more than an order of magnitude from those assumed. In a mean-field calculation $\chi_{s0}$ 
does not change above $T_c$. There is, however, a jump predicted in $d\chi/dT$ expected at $T_c$.

\section{Summary}\label{Summary}
We have studied the evolution of the specific heat and other thermodynamic properties in an effective 
theory of fluctuating orbital currents in high-$T_c$ cuprates. The motivation for the work has been 
to see if the finite-temperature break-up of a proposed ordering associated with a loop 
current pattern is consistent with both the existence of an order parameter in the pseudogap phase below 
a temperature $T^*(x)$, and with an {\it absence} of an observed singularity in the specific 
heat and a weak singular feature in the uniform magnetization at $T^*(x)$.  This is a first step towards 
investigating, through quantum Monte Carlo simulations, whether the quantum break-up of such order gives 
rise to quantum critical fluctuations that could possibly explain the anomalous transport properties in 
the normal state of these compounds, as has been proposed in analytic calculations \cite{Aji-Varma}. In 
this paper, we have shown that the effective field theory of the particular proposed  order of orbital 
currents within a $CuO_2$-plane passes this test by destroying the order while exhibiting no divergence 
in the specific heat. Instead, we have found bumps which we have estimated to be of a magnitude that
are unobservable in experiments done so far. Moreover, we find a  uniform magnetic susceptibility  with  
a non-analytic behavior as a function of temperature as the phase transition is crossed. From a technical 
point of view, a principal result of our calculations is that the anisotropy considered in the Ashkin-Teller 
model as well as the next nearest neighbor interactions, in the range of parameters considered, are irrelevant 
perturbations.

\indent \textit{Acknowledgments}. This work was supported by the Norwegian Research Council 
Grants No. 158518/431 and No. 158547/431 (NANOMAT), and Grant No. 167498/V30 (STORFORSK). 
The authors acknowledge  communications and discussions with V. Aji, K. B{\o}rkje, E. H. Hauge, 
B. Leridon, J. Linder, A. Shekhter, Z. Tesanovic, and M. Wallin.

\end{document}